\relax
\documentstyle[prl,aps,epsfig,twocolumn,graphics,rotate]{revtex}
\begin {document}

\draft
\title{ From band insulator to Mott insulator in one dimension}
\author{Michele Fabrizio$^{1,3}$, Alexander O. Gogolin$^2$
and Alexander A. Nersesyan$^{3,4}$}
%and nobody else}
\address{
$^1$International School for Advanced Studies and INFM, 
Via Beirut 4, 34014 Trieste, Italy\\ 
$^2$Department of Mathematics, Imperial College, 180 Queen's Gate,
London SW7 2BZ, United Kingdom\\
$^3$ The Abdus Salam International Centre for Theoretical Physics,
P.O.Box 586, 34100, Trieste, Italy\\
$^4$Institute of Physics, Tamarashvili 6, 380077, Tbilisi, Georgia
}
\date{Draft: \today}
\maketitle
\begin{abstract}
We derive the phase diagram for the one-dimensional model of a 
ferroelectric perovskite recently introduced by Egami, Ishihara and
Tachiki [{\sl Science}, {\bf 261}, 1307 (1993)]. 
We show that  
the interplay between covalency, ionicity and strong correlations 
results in a spontaneously dimerized phase which
separates the weak-coupling band insulator from the 
strong-coupling Mott insulator.
The transition from the band insulator to the 
dimerized phase is identified 
as an Ising critical point.
The charge gap vanishes at this single point with the
optical conductivity diverging as
$\sigma(\omega)\sim \omega^{-3/4}$.
The spin excitations are gapless above the second
transition to the Mott insulator phase.
\end{abstract}

Much effort have been devoted to understanding the response
of strongly correlated electron systems to lattice distortions. 
Generally, one expects that the tendency to a charge density wave 
instability modulated by a lattice distortion is increased by 
(repulsive) interactions.  
A surprisingly pronounced enhancement has recently been 
observed in numerical studies of a prototype one dimensional (1D) model for 
ferroelectric perovskites, 
proposed by Egami, Ishihara and Tachiki\cite{Egami}.

The Hamiltonian of their model is 
\begin{eqnarray}
\hat{H}& =& \sum_{i\sigma}\left[- t\left( 
c^\dagger_{i\sigma}c^{\phantom{\dagger}}_{i+1\sigma} + 
{\rm H.c.}\right) 
+ \Delta (-1)^i n_{i\sigma}\right]
\nonumber\\& +& 
U \sum_i n_{i\uparrow}n_{i,\downarrow}\, , 
\label{Ham}
\end{eqnarray}
at half-filling.
The odd and even sites represent oxygen atoms (O) 
and a generic cation (C),
respectively, with the energy difference $E_C - E_O = 2\Delta$.
The hopping amplitude $t$ 
determines the amount of covalency and $U$ stands for the on-site 
Coulomb repulsion. 
In spite of its apparent simplicity, model
(\ref{Ham}) reveals nontrivial physics. 
At $U=0$ it describes 
a band insulator (BI) of a mixed ionic-covalent character, 
with a spectral gap for all 
excitations. 
At $\Delta = 0$ it is a Mott insulator (MI) with a 
finite 
charge gap, $m_c$, 
but with gapless spin excitations.
In fact, in the strong-coupling limit
$U\gg (t,\Delta)$, (\ref{Ham}) can be  
mapped onto a Heisenberg model 
by projecting out all doubly occupied sites.  
At the leading order, only a nearest neighbour exchange 
$J = 4t^2U/(U^2-4\Delta^2)$ is generated. 
The resulting Heisenberg chain is well known to possess a
gapless spin spectrum. 
%and  power-law decaying spin correlation functions.
The issue of interest is the nature of the crossover from the 
BI regime to the MI regime which, on general grounds,
is expected to occur in the region
where the single-particle BI
gap $\Delta$ becomes comparable 
with the MI charge gap $m_c$. 

Thus, increasing $U$ at fixed $\Delta$, one expects a transition 
to a spin-gapless phase.
In addition to this {\it spin transition},
the charge degrees of freedom 
should be also affected in the course 
of the crossover. 
This can easily be understood in the 
case $t \ll (U, \Delta)$ where, upon increasing $U$,
the system approaches a 
mixed-valence regime in which the 
two charge configurations, O$^{-2}$C$^{+2}$ with  
energy $U-2\Delta$ and O$^{-1}$C$^{+1}$ with zero energy,  
become degenerate.
Therefore, 
those excitations responsible for the charge redistribution 
among these two configurations
should soften around some value of $U$. 
The finite-size simulations 
\onlinecite{Egami,Egami:numerics,Martin:numerics,Sandro,RSII}
suggest that the mixed-valence regime is accompanied by a strongly
enhanced 
response to the coupling with zone-centre optical phonons $\xi$,   
%\begin{equation}
%\hat{H}_{ep} = -\lambda\xi \sum_{i\sigma}
%(-1)^i \left( 
%c^\dagger_{i\sigma}c^{\phantom{\dagger}}_{i+1\sigma} + {\rm
%H.c.}\right)\equiv -\lambda \xi \sum_{i} {\cal D}_i,
%\label{H:ep}
%\end{equation}
\begin{equation}
\hat{H}_{ep} = -\lambda \xi \sum_{i} {\cal D}_i,\,
{\cal D}_i = \sum_{\sigma}
(-1)^i \left( 
c^\dagger_{i\sigma}c^{\phantom{\dagger}}_{i+1\sigma} + {\rm
H.c.}\right),
\label{H:ep}
\end{equation}
where $\lambda$ is the electron-phonon coupling constant and ${\cal
D}_i$
is the dimerization (or staggered bond-density) operator.
In fact, the so-called average dynamical charge 
$Z_* =lim_{\xi\to 0} (P(\xi)-P(0))/\xi$, where 
$P(\xi)$ is the polarization, diverges  
at a particular point, even in finite-size 
simulations\cite{Sandro,Martin:numerics}. 
This is due to the
{\sl accidental} degeneracy
between two singlet ground states with opposite parity
which occurs in any finite-size system with the
number of sites multiple of 4\cite{Sandro,Martin:numerics}. 
More importantly, Resta and Sorella\cite{RSII} have recently observed
that the localization
length of the ground-state wave function 
diverges at some value of $U$,  which 
suggests vanishing of the charge gap; hence the {\it charge
transition}.

However, the overall picture which emerges from numerical 
simulations and the existing mean-field calculations
\cite{Martin:numerics},
clearly
underestimating the crucial role of fluctuations in 1D,
is still far from being satisfactory. 
It remains unclear whether the charge instability occurs at the 
same point as the spin transition.
Even the nature of the transition (transitions?) is not known. 
Is it a first-order transition or is it a second-order one? 
In the latter case, what is the universality class it belongs to? 
To the best of our knowledge, there have been no consistent analytic
attempts to resolve these intriguing questions.

In this Letter we present theoretical arguments
showing that there are two continuous transitions:
the spin transition of the Kosterlitz-Thouless (KT) type 
at $U = U_{c2}$, 
and the charge
transition at $U = U_{c1} < U_{c2}$, 
the latter identified as an Ising
critical point where the charge gap vanishes. In the 
intermediate region, $U_{c1} < U < U_{c2}$, 
the site-parity is spontaneously broken, and
the system is 
characterised by a doubly degenerate, dimerized ground state,
with $\langle {\cal D}_i \rangle$ being the order parameter. 

We start our discussion with addressing the 
mechanism of the spin gap generation in the MI phase.
For the single-chain Hamiltonian (\ref{Ham}),  
which is SU(2) and site-parity invariant, the 
only possibility
compatible with the symmetries is a  
spontaneous spin-dimerization, like the one occurring 
at $J_2 / J_1 \simeq 0.24$ in the 
frustrated Heisenberg chain 
with nearest-neighbour ($J_1$) and next-nearest-neighbour ($J_2$)
exchanges. 
The same scenario is realized in 
our case below a 
critical value $U_{c2}$ which depends on $\Delta$ and $t$. 
This result  
agrees with the exact diagonalization of 
a 4-site chain where the  
next-to-nearest neighbour exchange starts to play a role.
At large $U$, the ground state is an odd-parity singlet, and 
the lowest excitation is a spin-triplet, followed at higher energy by
an 
even-parity singlet. 
As $U$ decreases, first the even-parity singlet 
and the triplet cross, and finally the two opposite-parity singlets 
cross. 
For $(U,\Delta) \gg t$ these two level crossings occur at 
$U\simeq 2\,\Delta + 1.86\, t$ and at $U\simeq 2\,\Delta + t$,
respectively (see also Ref.\cite{Martin:numerics}). 
The same  
is observed in the 
$J_1$-$J_2$ Heisenberg model when $J_2$ is increased \cite{J1J2}. 
The crossing of the triplet and singlet excited states 
in finite-size systems becomes 
the spontaneous spin-dimerization transition in the thermodynamic
limit. 
The crossing of the two opposite-parity singlets,
taking place at $J_2=0.5\, J_1$, corresponds to the Majumdar-Gosh
solvable 
limit\cite{Majumdar}.

The appearance of a spontaneously 
dimerized insulating (SDI) phase can be also inferred
starting from the BI phase. In this limit we need an 
improved version of the mean-field calculation presented in 
Ref.\cite{Martin:numerics}.  
We introduce a two component Fermi field
$
\Psi^\dagger_{k\sigma} = 
( c^\dagger_{k\sigma}\, ,\, c^\dagger_{k+\pi\sigma} ),
$
$~k\in [-\pi/2,\pi,2]$, and the operators
\begin{equation}
\hat{\tau}_{a\sigma}(q) = \sum_{k}
\Psi^\dagger_{k\sigma} \tau_a \Psi^{\phantom{\dagger}}_{k+q\sigma},
~~(a=0,1,2,3),
\label{tau}
\end{equation}
where $\tau_a$ are the Pauli matrices including
the unit matrix $\tau_0$. 
The staggered charge/spin densities correspond to
$\hat{\tau}_{1\uparrow}(0) \pm \hat{\tau}_{1\downarrow}(0)$,
while the dimerization is  $\hat{\tau}_{2\uparrow}(0) + 
\hat{\tau}_{2\downarrow}(0)\equiv \sum_i {\cal D}_i$
[see Eq.(\ref{H:ep})].
A generic interaction 
compatible with the symmetry of (\ref{Ham}) can be written as 
\begin{eqnarray}
&&\hat{H}_{int} = \frac{1}{4L}\sum_{q,\sigma,\sigma'} [
(g_1+g_3) \hat{\tau}_{1\sigma}(q)\hat{\tau}_{1\sigma'}(-q) 
\nonumber\\
&\mbox{}& + (g_1-g_3)
 \hat{\tau}_{2\sigma}(q)\hat{\tau}_{2\sigma'}(-q) +
2 g_2\, \hat{\tau}_{0\sigma}(q)\hat{\tau}_{0\sigma'}(-q) ]. 
\label{Hint}
\end{eqnarray}
The coupling constants are written using the $g$-ology jargon\cite{Solyom}: 
$g_1, g_2$ and $g_3$ label
the backward, forward and Umklapp scattering amplitudes, respectively.
In the special case of (\ref{Ham}), 
$g_1=g_2=g_3=U$, and the $\hat{\tau}_2$-$\hat{\tau}_2$ term vanishes.  
Therefore, in any mean-field treatment of the  
original Hamiltonian, 
only variational states with finite average values of 
$\hat{\tau}_{1\sigma}(0)$, i.e. the staggered charge and spin 
densities, can be explored. 
However, the equality among the $g$'s is already
lost at the first order of perturbation theory. More refined weak-coupling
schemes, 
such as renormalization group\cite{Solyom} or the bosonization
technique\cite{GNT}, show that renormalization occurring at energies well below
the bandwidth cutoff but still larger than $max (\Delta, m_c)$
enhances $g_3$ and $g_2$ but suppresses $g_1$.
This implies generation of an effective coupling
$(g_3-g_1)$ which may lead to 
the appearance of a finite average value of $\hat{\tau}_2(0)$, in
addition to the average
$\langle \hat{\tau}_1(0) \rangle$ induced by the $\Delta$-term.
By including these renormalization effects within the leading
logarithmic 
approximation\cite{long}, 
we indeed find a sequence of two transitions.
First, at $U=U_{c1}$, a finite 
average $\langle \hat{\tau}_2(0) \rangle $ continuously appears.
At a larger $U=U_{c2}$
a triplet exciton gets soft signalling 
the transition to the MI phase. Although these two transitions 
are physically distinct within this scheme, to estimate the difference 
between $U_{c1}$ and $U_{c2}$ 
one has to go beyond the leading logarithmic approximation, which is
not feasible. For this reason, we can only safely state that
$U_{c1}<U_{c2}$, 
with both critical points being of the order $2\pi t /\ln(t/\Delta)$.

In what follows, 
we develop the {\it low-energy} effective field theory
for the lattice model (\ref{Ham}). 
Considering the
weak-coupling case, $(U,\Delta) \ll t$, we 
linearise the spectrum and pass to the continuum limit by
substituting
$
a_0^{-1/2}c_{n\sigma}\rightarrow 
i^n \psi_{R\sigma}(x)+(-i)^n \psi_{L\sigma}(x)$, $x=na_0$, 
where $a_0$ is the lattice spacing, and $\psi_{R,L} (x)$
are the right and left components of the Fermi field. 
These fields can be bosonized in a standard way\cite{GNT}: 
$
\psi_{R,L;\sigma} = (2\pi a_0)^{-1/2}
e^{\pm i\sqrt{4\pi}\phi_{R,L;\sigma}}
$, where $\phi_{R(L),\sigma}$ are the right(left)-moving Bose
fields. We define $\Phi_\sigma=\phi_{R\sigma}+
\phi_{L\sigma}$ and introduce linear combinations,
$\Phi_c=(\Phi_\uparrow+\Phi_\downarrow)\sqrt{2}$ and
$\Phi_s=(\Phi_\uparrow-\Phi_\downarrow)\sqrt{2}$, 
to describe the charge and spin degrees
of freedom, respectively. 
Then the Hamiltonian density of the bosonized model is given by:
$
{\cal H}_{\em eff} = {\cal H}_c + {\cal H}_s
+ {\cal H}_{cs}.
$
Here the charge and spin sectors are described by
\begin{eqnarray}
{\cal H}_c &=& \frac{v_c}{2}\left[ \Pi_c^2  + 
(\partial_x \Phi_c)^2\right]
\nonumber\\
 &-& \frac{g_3}{\pi^2 a_0}\cos\sqrt{8\pi}\Phi_c
+ \frac{2 g_2 a_0}{\pi} \partial_x \phi_{cR}\partial_x \phi_{cL}
\label{charge-ham} \\
{\cal H}_s &=& \frac{v_s}{2}\left[ \Pi_s^2  +(\partial_x\Phi_s)^2\right]
\nonumber\\
 &+& \frac{g_1}{\pi^2 a_0}\cos\sqrt{8\pi}\Phi_s
- \frac{2 g_1 a_0}{\pi} \partial_x \phi_{sR}\partial_x \phi_{sL}
\label{spin-ham}
\end{eqnarray}
where $\Pi_{c,s}$ are the momenta conjugate to $\Phi_{c,s}$, and
$v_{c,s}$ are the velocities of the charge and spin excitations.
Actually, ${\cal H}_s$ in (\ref{spin-ham}) is an Abelian bosonized
version of the SU(2)$_1$-symmetric Wess-Zumino-Novikov-Witten model
with a marginally irrelevant ($g_1 > 0$) current-current perturbation,
$- g_1 {\bf J}_R \cdot {\bf J}_L$, given by the last two terms in
(\ref{spin-ham}).  
The two sectors of the theory are coupled by the $\Delta$-term:
\begin{equation}
{\cal H}_{cs} = - (2\Delta /\pi a_0) \sin\sqrt{2\pi}\Phi_c
\cos \sqrt{2\pi} \Phi_s .
\label{cs-coupling}
\end{equation}
In fact, the effective Hamiltonian ${\cal H}_{\em eff}$
represents the low-energy limit for a wide class of lattice
models including (\ref{Ham}).
Such lattice effects as nearest-neighbour repulsion, or 
a difference between the Coulomb repulsion on the O-atoms and the
C-atoms,
will only renormalize the parameters of ${\cal H}_{\em eff}$,
in particular the charge stiffness constant ($K_c$) and
the velocities $v_{c,s}$.

Starting from the MI phase, 
one can firmly justify the opening of a spin
gap and spontaneous dimerization upon decreasing $U$.
Assuming $U \gg \Delta$, 
in which case the spin sector is gapless
while the charge one is gapped, 
one can integrate out the charge degrees
of freedom to obtain at the second order in $\Delta$ a nontrivial
renormalization of $g_1$: 
$
\tilde{g}_1 = g_1 - \left( \Delta /m_c \right)^2 \left( v_c / a_0
\right).
$
Until $\tilde{g}_1 > 0$, the spin spectrum remains gapless.
However,  when $U$ is decreased, 
$\tilde{g}_1$ eventually becomes negative, 
and the system undergoes a continuous (KT) transition 
to the dimerized phase with a nonzero spin gap.
Using the exact result for the small-$U$ Hubbard model, 
$
m_c\sim \sqrt{Ut}e^{-2\pi t/U}
$, 
one concludes that
\begin{equation}
U_{c2}= \frac{2 \pi t}{\ln (t/\Delta)}\left[ 1 +
O \left( \frac{\ln \ln (t/\Delta)}{\ln (t/\Delta)}\right) \right],
%\label{Uc2}
\end{equation}
in agreement with the previous estimate.

Bosonization also offers a simple understanding of the region
$U<U_{c2}$. 
In what follows, we shall focus on the nature of the 
transition at $U =U_{c1}$. To this end, 
the Hamiltonian
${\cal H}_{\em eff}$, given by (\ref{charge-ham})-(\ref{cs-coupling}),
will be regarded 
as a {\sl phenomenological}
Landau-Ginzburg energy functional, in the sense that all  
the couplings are effective ones obtained by integrating out 
high-energy degrees of freedom. 

We start with looking
at the saddle points of the effective potential in
${\cal H}_{\em eff}$ which obey the equations
(here $\varphi_{c,s} = \sqrt{2\pi} \Phi_{c,s}$)
\begin{eqnarray}
\cos \varphi_c\left
[4\tilde{g}_3\sin\varphi_c - 2\pi \tilde{\Delta} 
\cos\varphi_s\right]&=&0 \label{phic}\\
\sin\varphi_s\left[ 2\pi \tilde{\Delta} 
\sin\varphi_c - 4\tilde{g}_1\cos\varphi_s\right]&=& 0, 
\label{phis}
\end{eqnarray}
supplemented by 
stability conditions.
At $\tilde{g}_3< \pi \tilde{\Delta}/2$ one finds two sets 
of minima (defined modulo $2\pi$) 
located at  
$\varphi_s = 0$, $\varphi_c = \pi/2$ ,
and 
$\varphi_s = \pi $, $\varphi_c = -\pi/2$. 
These sets characterise the BI phase. 
Indeed, the vacuum-vacuum 
transitions, $\Delta \varphi_{s(c)} = \pm \pi$,
describe stable topological excitations carrying the charge 
$Q=\Delta \varphi_c /\pi =\pm 1$
and spin $S^z =\Delta \varphi_s / 2\pi = \pm 1/2$ 
and therefore coinciding with ``massive''
single-fermion excitations of the BI.
The situation changes at $\tilde{g}_3> \pi \tilde{\Delta}/2$ where 
each minimum in the charge sector splits into two 
degenerate minima and thus
transforms to a local {\sl double-well} potential.
The new minima are given by
$\varphi_s = 0$, $\varphi_c = \phi_0$,  $\pi -\phi_0$,
and 
$\varphi_s = \pi$, $\varphi_c = -\phi_0$, $-\pi +\phi_0$,
where 
$
\phi_0 = \arcsin (\pi \tilde{\Delta}/2 \tilde{g}_3)$. 
They describe the SDI phase.
In this region, the dimerization operator
$
{\cal D} (x) \sim \cos \varphi_c (x) \cos \varphi_s (x)
$
acquires a finite expectation value.
Notice that the location of the minima in the spin sector,
and hence the spin quantum numbers of the topological excitations,
are the same as in the BI phase
($\tilde{g}_3< \pi \tilde{\Delta}/2$). 
Therefore the spin part of the spectrum formed in the SDI phase
smoothly transforms into that of the BI phase. The transition
at
$\tilde{g}_3= \pi \tilde{\Delta}/2$  
mainly involves the charge degrees of freedom
which undergo dramatic changes.
In particular,  the charge quantum numbers become
{\sl fractional}, depending on $\phi_0$.
The $Z_2$-degeneracy of the SDI state implies the existence of 
topological kinks carrying the spin S = 1/2
and charge  $Q = \pm 2\varphi_0 /2$. These massive excitations,
described by interset vacuum-vacuum transition, 
interpolate between the neutral spinons of the MI phase
and single-fermion excitations of the BI phase.
However, 
the double-well local structure of the effective potential in the
charge sector gives rise to {\sl singlet} kink excitations as well, 
whose mass and charge $Q = 1-2\phi_0/\pi$
vanish at $\tilde{g}_3= \pi \tilde{\Delta}/2$. 
On the BI side of the phase diagram, these singlet excitations appear
as excitonic bound states inside the single-particle gap.
Precisely at  $\tilde{g}_3= \pi \tilde{\Delta}/2$,
the effective potential becomes $\sim \varphi^4 _c$,
which is well known to describe the Ising universality class.

We have checked that the SDI solution remains stable in the region
\begin{equation}
\sqrt{\tilde{g}_1 \tilde{g}_3} \leq \frac{\pi}{2}\tilde{\Delta} 
\leq \tilde{g}_3.
\end{equation}
This condition can be satisfied if $\tilde{g}_1 < \tilde{g}_3$, 
which is indeed true for our phenomenological Hamiltonian.
However, separation between 
the transition points $U_{c1}$ and $U_{c2}$
becomes even stronger when the electron-phonon interaction,
always present in realistic situations, is taken into account.
Let us assume that the zone-centre optical phonons are
dispersionless, with a characteristic frequency
$\omega_0$ being the largest energy scale in the problem. In this
limit, the phonons 
mediate an instantaneous  
interaction between the
electrons which adds to the Hubbard repulsion and leads to a
renormalization of the ``bare'' coupling constants:
$g_3 \rightarrow g_3 + g_0$, $g_1 \rightarrow g_1 - g_0$,
where $g_0 \sim \lambda^2/\omega_0$. 
The transition points $U_{c1}$ and $U_{c2}$
will start moving apart
upon increasing $g_0$, and if the electron-phonon coupling
is large enough ($g_0 > g_1$), a finite spin gap will be generated
even at $\Delta = 0$. 
Qualitatively the same situation occurs upon 
relaxing the condition of 
extremely quantum phonons (large $\omega_0$)
at a fixed $\lambda$.
This observation explains that the two transitions
can be, in principle, completely disentangled and, on the other hand, 
it provides an extra support to the fact that the charge transition
separates the BI and SDI phases.

The Ising scenario at $U = U_{c1}$
can be more rigorously proven by 
studying the charge degrees of freedom with 
the spin bosonic field locked at $\Phi_s = 0$ or $\sqrt{\pi/2}$.
Such approach will certainly be valid 
in a small vicinity of $U_{c1}$
because, as we already mentioned, the structure of the gapped spin
sector is qualitatively the same both in the BI and SDI phases.
The charge sector can then be 
described by 
a double sine-Gordon Hamiltonian 
\begin{eqnarray}
&&\hat{H}_c(x) = \frac{v_c}{2}\left[\Pi_c(x)^2 + 
\left(\partial \Phi_c(x)\right)^2 \right] \nonumber\\
&-& \frac{g}{\pi^2 a_0}\cos\sqrt{8\pi K_c}\Phi_c(x) 
- \frac{2\Delta}{\pi a_0}
\sin\sqrt{2\pi K_c}\Phi_c(x),
\label{Hcharge-eff} 
\end{eqnarray}
where, we stress once more, all the parameters are 
purely phenomenological, and 
part of the interaction has been 
absorbed in the rescaling parameter $K_c<1$.
A quantum phase transition of an Ising type in 
the double sine-Gordon model was recently discovered by 
Delfino and Mussardo\cite{D&M}.
In the whole parameter region of interest to us, i.e. 
$g>0$ and $K_c<1$, they argued that 
the existence of an Ising critical point is a universal property of
this model. This allows us to investigate this model 
in the vicinity of the point 
$K_c=1/2$, where the problem greatly simplifies.
At this point, the Hamiltonian (\ref{Hcharge-eff}) can be mapped 
onto two coupled 2D Ising 
models, or equivalently, two coupled quantum 
Ising chains in a transverse magnetic field
(a deformation of the Ashkin-Teller model).
One of the two chains  always remains off-critical, 
while the other one can pass through a 
critical point by fine tuning the parameters of the model. 
The details will 
be presented elsewhere\cite{long}. 
Here we just quote our main findings,
which can be easily understood once the Ising scenario is assumed. 
The most relevant perturbation at an Ising transition 
is the magnetic field $h$,  
which, in the ordered phase, selects one of the two equivalent vacua. 
By analogy, the role of $h$ is played in our case by the explicit 
dimerization $\xi$,  Eq.(\ref{H:ep}).
The dimerization operator ${\cal D}$
(which also plays the role of the charge polarizaton
operator) is therefore
proportional to the order parameter $\sigma$ of
the quantum critical Ising model. 
The Ising mapping also allows us to identify
the electron current operator, $j\sim \partial_t \sigma$, 
and the average electron charge operator 
$\rho\sim - \partial_x \sigma$.
These findings are perfectly consistent with
the general relations 
$\rho\sim -\partial_x{\cal D}$ and
$\partial_t\rho+\partial_x j=0$ (continuity) valid for
insulators (i.e., in the absence of free carriers).
>From known results on the correlation functions at the 
Ising transition 
we conclude that the susceptibility to  
the explicit dimerization, namely the real part of the 
polarizability $\chi(\omega)$, diverges at low frequencies as
$\omega^{-7/4}$, and  
the optical conductivity, 
$\sigma(\omega)\sim \omega^{-3/4}$ (no Drude peak),
is `semi-metallic'.

In conclusion, we have identified the origin of the anomalous
response to an electron-phonon coupling in the 1D ferroelectric
perovskite
model (\ref{Ham}), which turns out to be far
stronger than that of free electrons ($\Delta=U=0$), 
of a band insulator ($U=0$),
or of a pure Hubbard model ($\Delta=0$).  
Our findings are summarised in Fig.1,  
where the phase diagram is shown at 
fixed $\Delta$ and $t$ as a function of $U$.  
Between the BI  ($U<U_{c1}$) and MI ($U>U_{c2}$) phases 
there exists a  
spontaneously dimerized phase 
even in the 
absence of an explicit electron-phonon interaction\cite{Gidopoulos}.
The transition 
from the MI to the spontaneously dimerized insulator
is of the KT type, accompanied by 
opening of the spin gap.
The transition 
from the dimerized phase to the BI has been shown
to be of the 
Ising type, with the charge gap   
closing at this single point\cite{Gidopoulos}.
   
We acknowledge useful discussions with 
D. Edwards, Yu Lu, and E. Tosatti. We 
particularly thank G. Mussardo and S. Sorella for their enlightening
comments. 
MF is partly supported by INFM, under project PRA HTSC.  
AOG is supported by the EPSRC of the United Kingdom.

%%%%%%%%%%%%%%%%%%%%%%%%%%%%%%%%%%%%%%%%%%%%%%%%%

\begin{figure}
\centerline{\epsfig{file=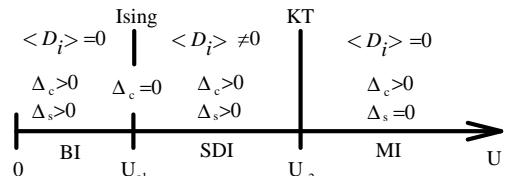}}
\caption{Phase diagram of (\protect\ref{Ham}\protect) as function 
of $U$ at fixed $\Delta$ and $t$. $\Delta_{c(s)}$ stands for the 
charge(spin) gap, ${\cal D}_i$ is the dimerization operator, whose 
average is finite in the spontaneously dimerized insulator  
but vanishes both in the band and in the Mott insulators. }
\end{figure} 

\end{document}